\documentclass{ifacconf}

\usepackage{amsmath}
\usepackage{amssymb}
\usepackage{upgreek}
\usepackage{algorithm}
\usepackage{algpseudocode}
\usepackage{verbatim}

\usepackage{graphicx}      
\usepackage{natbib}        

\usepackage{booktabs,caption}
\usepackage{units}
\usepackage{subfig}
\usepackage{empheq}
\usepackage{cancel}
\usepackage{enumitem}
\usepackage{epstopdf}
\usepackage{url}

\usepackage{tikz}
\usepackage{pgfplots}
\pgfplotsset{compat=newest}
\usepackage{pgfkeys}
\newlength\myheight
\newlength\mywidth
\usetikzlibrary{plotmarks}
\usetikzlibrary{shapes,arrows,shadows,positioning}
\usepackage{tikzscale}
\usetikzlibrary{decorations.pathreplacing}
\usetikzlibrary{shapes,arrows,patterns}

\newtheorem{theorem}{Theorem}
\newtheorem{corollary}{Corollary}

\newtheorem{definition}{Definition}
\newtheorem{remark}{Remark}

\newcommand{\Summationsstelle}[4][]{ %
  \begin{scope}[thick, #1]
    \node[draw, circle, minimum width = #4, minimum height = #4] (#2) at (#3) {};
  \end{scope} %
}

\newcommand{\Verzweigung}[4][]{ %
  \begin{scope}[inner sep = 0 pt, outer sep = 0 pt, thick, #1]
    \node[circle, minimum width = #4, minimum height = #4] (#2) at (#3) {};
    \draw[solid, fill] (#2) circle (0.5*#4);
  \end{scope} %
}

\newcommand{\UeFunk}[5][]{ %
  \begin{scope}[thick, #1]
    \node[minimum width = #4, minimum height = #4] (#2) at (#3) {#5};
    \draw (#2.north east) rectangle (#2.south west);
  \end{scope} %
}

\begin{document}
\begin{frontmatter}

\title{Nonlinear Moment Matching for the Simulation-Free Reduction \\ of Structural Systems} 


\author{Maria Cruz Varona, Nico Schneucker, Boris Lohmann} 

\address{Chair of Automatic Control, Technical University of Munich, Boltzmannstr. 15, D-85748 Garching, Germany \\ (email: maria.cruz@tum.de).}

\begin{abstract}                
This paper transfers the concept of moment matching to nonlinear structural systems and further provides a simulation-free reduction scheme for such nonlinear second-order models. \\After first presenting the steady-state interpretation of linear moment matching, we then extend this reduction concept to the nonlinear second-order case based on~\cite{astolfi2010model}. Then, similar simplifications as in \cite{cruz2019practicable} are proposed to achieve a simulation-free nonlinear moment matching algorithm. A discussion on the simplifications and their limitations is presented, as well as a numerical example which illustrates the efficiency of the algorithm.
%
\end{abstract}


\end{frontmatter}

\vspace{-0.5em}
\section{Introduction} 
\vspace{-1em}

Model order reduction (MOR) for nonlinear structural dynamics has gained an increased importance during the last years. This is due to the large number of engineering applications, where reduced order models are indispensable for the efficient analysis and computer-aided optimization of mechanical structures undergoing large deformations, and/or exhibiting nonlinear material behavior. In this regard, two different reduction streams can be distinguished.

  

Simulation-based \emph{dimensional reduction} techniques such as Proper Orthogonal Decomposition (POD), often combined with \emph{hyper reduction} methods for the efficient evaluation of the nonlinear terms via e.g. the Energy-Conserving Sampling and Weighting (ECSW), are established and successful nonlinear approaches (see e.g. \cite{farhat2014dimensional}). 
Known in the context of nonlinear frequency response/mo\-dal analysis (\cite{kerschen2009nonlinear}), the Harmonic Balance Method (HBM) and the Nonlinear Normal Modes (NNMs) have lately become very popular also in the model reduction field (\cite{weeger2014nonlinear,sombroek2018numerical}). These procedures, however, rely on either numerical continuation methods or shooting techniques, making them also fairly computational expensive in comparison to a numerical integration needed e.g. for POD.
\\
Simulation-free reduction procedures have been also studied in the last years. In particular, the concept of basis augmentation with so-called modal derivatives (MDs) has been successfully applied as reduction technique in various structural applications (see e.g. \cite{rutzmoser2018model} and references therein). The key idea is to first compute some dominant vibration modes of the linearized, second-order system, and then to augment the reduction basis with perturbation derivatives capturing the nonlinear behavior.

A few years ago, the concept of moment matching and Krylov subspaces known from linear MOR has been carried over to nonlinear \emph{first-order} systems by \cite{astolfi2010model}. The extension is based on the steady-state interpretation of linear moment matching, as well as on the steady-state res\-ponse of nonlinear systems, the center manifold theory and the techniques of nonlinear output regulation \cite[Ch. 8]{isidori1995nonlinear}, \cite{huang2004nonlinear}. Although the method is system-theoretically attractive, it involves the difficult solution of a nonlinear partial differential equation (PDE). Thus, some simplifications have been proposed recently towards a \emph{practicable, simulation-free} nonlinear moment matching algorithm (\cite{cruz2019practicable}). The proposed \emph{reduction approach} is related to the data-driven, low-order nonlinear system \emph{identification technique} from \cite{scarciotti2017data}, in the sense that both methods approximately match nonlinear moments.

The goal of this paper is to transfer the concept of moment matching for linear structural systems to the nonlinear \emph{second-order} case. In Section~\ref{sec:LMM} we first present the time domain (aka. steady-state) interpretation of moment matching for linear second-order systems. Based on \cite{astolfi2010model}, we then transfer this reduction method to the nonlinear second-order case and provide the corresponding nonlinear, Sylvester-like PDE. In Section~\ref{sec:simulation-free-NLMM}, simplifications are proposed to avoid the PDE, and achieve a simulation-free second-order nonlinear moment matching algorithm, which relies on the solution of nonlinear systems of equations. Finally, numerical results for a geometrically nonlinear structural model are provided.



\vspace{-0.5em}
\section{Moment Matching for linear structural systems} \label{sec:LMM}
\vspace{-0.5em}
Consider a large-scale, linear time-invariant (LTI), asymptotically stable, multiple-input multiple-output (MIMO) second-order model of the form
\vspace{-0.2em}
\begin{align} \label{eq:linear-FOM}
\boldsymbol{M} \ddot{\boldsymbol{q}}(t) + \boldsymbol{D} \dot{\boldsymbol{q}}(t) + \boldsymbol{K} \boldsymbol{q}(t) &= \boldsymbol{B} \boldsymbol{F}(t), \ \boldsymbol{q}(0) \!=\! \boldsymbol{q}_0, \ \dot{\boldsymbol{q}}(0) \!=\! \dot{\boldsymbol{q}}_0 \nonumber \\
\boldsymbol{y}(t) &= \boldsymbol{C} \boldsymbol{q}(t),
\end{align}
where $\boldsymbol{M}\!, \boldsymbol{D}\!, \boldsymbol{K} \!\in\! \mathbb{R}^{n \times n}$ are the mass, damping and stiffness matrices, and $\boldsymbol{q}(t) \!\in\! \mathbb{R}^n$, $\boldsymbol{F}(t) \!\in\! \mathbb{R}^m$, $\boldsymbol{y}(t) \in \mathbb{R}^p$ denote the displacements, forces (inputs) and outputs of the system.
The goal of model reduction is to approximate the full order model (FOM) \eqref{eq:linear-FOM} by a reduced order model (ROM)
\begin{equation} \label{eq:linear-ROM}
\begin{aligned}
\boldsymbol{M}_{\mathrm{r}} \ddot{\boldsymbol{q}}_{\mathrm{r}}(t) + \boldsymbol{D}_{\mathrm{r}} \dot{\boldsymbol{q}}_{\mathrm{r}}(t) + \boldsymbol{K}_{\mathrm{r}} \boldsymbol{q}_{\mathrm{r}}(t) &= \boldsymbol{B}_{\mathrm{r}} \boldsymbol{F}(t), \\
\boldsymbol{y}_{\mathrm{r}}(t) &= \boldsymbol{C}_{\mathrm{r}} \boldsymbol{q}_{\mathrm{r}}(t),
\end{aligned}
\end{equation} 
of much smaller dimension $r \ll n$ with reduced matrices $\left\{\boldsymbol{M}_{\mathrm{r}}, \boldsymbol{D}_{\mathrm{r}}, \boldsymbol{K}_{\mathrm{r}}\right\} \!=\! \boldsymbol{W}^{\mathsf T} \left\{\boldsymbol{M}, \boldsymbol{D}, \boldsymbol{K}\right\} \boldsymbol{V}$, $\boldsymbol{B}_{\mathrm{r}} \!=\! \boldsymbol{W}^{\mathsf T} \boldsymbol{B}$, $\boldsymbol{C}_{\mathrm{r}} \!=\! \boldsymbol{C} \, \boldsymbol{V}$ and initial conditions $\boldsymbol{q}_{\mathrm{r}}(0) \!=\! \boldsymbol{W}^{\mathsf T} \, \boldsymbol{q}_0, \dot{\boldsymbol{q}}_{\mathrm{r}}(0) \!=\! \boldsymbol{W}^{\mathsf T} \, \dot{\boldsymbol{q}}_0$, such that $\boldsymbol{y}(t) \approx \boldsymbol{y}_{\mathrm{r}}(t)$. In this projection-based framework, the main task consists in finding suitable (orthogonal) projection matrices $\boldsymbol{V}, \boldsymbol{W} \in \mathbb{R}^{n \times r}$ that span appropriate subspaces. 
\begin{remark}
	Note that, for second-order systems, the reduction is usually performed by a \emph{Galerkin projection} with $\boldsymbol{W} \!=\! \boldsymbol{V}$ rather than by a two-sided (oblique) Petrov-Galerkin projection. This choice preserves the symmetry and definiteness of the original matrices, which implies the preservation of the stability and passivity of the FOM.
\end{remark}

\vspace{-0.7em}
\subsection{Notion of Moments and Krylov subspaces}
\vspace{-0.5em}

The transfer function of the system \eqref{eq:linear-FOM} is
\begin{equation}
\boldsymbol{G}(s) = \boldsymbol{C}(s^2 \boldsymbol{M} + s \boldsymbol{D} + \boldsymbol{K})^{-1} \boldsymbol{B}. 
\end{equation}
\begin{definition}
	The \emph{moments} $\boldsymbol{m}_i(\sigma)$ of $\boldsymbol{G}(s)$ at the complex expansion point $\sigma \!\in\! \mathbb{C}$ are given by
	\begin{equation*}
	\begin{aligned}
	\boldsymbol{m}_i(\sigma) 
	&= (-1)^i \, \left[\boldsymbol{C} \ \ \mathbf{0}\right]\begin{bmatrix}
	\boldsymbol{K}_{\sigma}^{-1} \boldsymbol{D}_{\sigma} & \boldsymbol{K}_{\sigma}^{-1} \boldsymbol{M}\\
	-\mathbf{I} & \mathbf{0}
	\end{bmatrix}^i \begin{bmatrix}
	\boldsymbol{K}_{\sigma}^{-1} \boldsymbol{B} \\ \mathbf{0}
	\end{bmatrix},
	\end{aligned}
	\end{equation*}
	where $\boldsymbol{K}_{\sigma} \!=\! \boldsymbol{K} + \sigma \boldsymbol{D} + \sigma^2 \boldsymbol{M}$ and $\boldsymbol{D}_{\sigma} \!=\! \boldsymbol{D} + 2\sigma \boldsymbol{M}$. 
\end{definition}
Depending on the considered damping, two different Krylov subspaces for second-order systems can be distinguished to achieve implicit moment matching:
\begin{itemize}[leftmargin=*]
	\item For general damping ($\boldsymbol{D} \neq \mathbf{0}$), so-called second-order Krylov subspaces
	\begin{subequations} \label{eq:2nd-multimom-block-Krylov}
		\begin{equation}
		\mathcal{K}_r^{\text{2nd}}\left(\boldsymbol{K}_{\sigma}^{-1} \boldsymbol{D}_{\sigma}, \boldsymbol{K}_{\sigma}^{-1} \boldsymbol{M}, \boldsymbol{K}_{\sigma}^{-1} \boldsymbol{B}\right) \subseteq \mathrm{ran}(\boldsymbol{V}),
		\end{equation}
		\begin{equation}
		\mathcal{K}_r^{\text{2nd}}\left(\boldsymbol{K}_{\mu}^{- \mathsf{T}}\boldsymbol{D}_{\mu}^{\mathsf T}, \boldsymbol{K}_{\mu}^{-\mathsf T} \boldsymbol{M}^{\mathsf T}, \boldsymbol{K}_{\mu}^{-\mathsf T}\boldsymbol{C}^{\mathsf T}\right) \subseteq \mathrm{ran}(\boldsymbol{W}),
		\end{equation}
	\end{subequations}
	are employed.
	These Krylov subspaces yield a two-stage Arnoldi-like recurrence, aka. second-order Arnoldi (SOAR). Further details concerning this case are available in \cite{bai2005dimension, salimbahrami2005structure}. 
	\item For proportional ($\boldsymbol{D} \!=\! \alpha \boldsymbol{M} + \beta \boldsymbol{K}$) or zero ($\boldsymbol{D} \!=\! \mathbf{0}$) damping, the classical first-order Krylov subspaces
	\begin{subequations} \label{eq:multimom-block-Krylov}
		\begin{align}
		\mathcal{K}_r^{\text{1st}}\left(\boldsymbol{K}_{\sigma}^{-1} \boldsymbol{M}, \boldsymbol{K}_{\sigma}^{-1} \boldsymbol{B}\right) \subseteq \mathrm{ran}(\boldsymbol{V}), \label{eq:multimom-block-Krylov-V} \\[0.2em] 
		\mathcal{K}_r^{\text{1st}}\left(\boldsymbol{K}_{\mu}^{-\mathsf T} \boldsymbol{M}^{\mathsf T}, \boldsymbol{K}_{\mu}^{-\mathsf T}\boldsymbol{C}^{\mathsf T}\right) \subseteq \mathrm{ran}(\boldsymbol{W}),
		\end{align}
	\end{subequations}
	can be employed instead, yielding -- exemplarily for $\boldsymbol{V} \!=\!~\left[\boldsymbol{V}_0, \ldots, \boldsymbol{V}_{r-1}\right]$ -- the one-stage Arnoldi-like recurrence
	\begin{equation}
	\begin{aligned}
	(\boldsymbol{K} + \sigma \boldsymbol{D} + \sigma^2 \boldsymbol{M}) \, \boldsymbol{V}_0 &= \boldsymbol{B}, \\
	(\boldsymbol{K} + \sigma \boldsymbol{D} + \sigma^2 \boldsymbol{M}) \, \boldsymbol{V}_i &= \boldsymbol{M} \, \boldsymbol{V}_{i-1}, \quad i \geq 1. 
	\end{aligned}
	\end{equation}
	For further details, see \cite{beattie2005krylov,salimbahrami2005structure}. 
\end{itemize}
Note that, in addition to the \emph{multimoment} matching strategy, it is also possible to match (high-order) moments at a set of different shifts $\left\{\sigma_i\right\}_{i=1}^q$ and $\left\{\mu_i\right\}_{i=1}^q$ with associated multiplicities $\left\{r_i\right\}_{i=1}^q$ (aka. \emph{multipoint} moment matching). Also note that, besides the \emph{block} Krylov subspaces, in the MIMO case we alternatively may use so-called \emph{tangential} Krylov subspaces (e.g. for~\eqref{eq:multimom-block-Krylov} and~$r_1=\ldots=r_q = 1$):
\begin{subequations} \label{eq:multipoint-tang-Krylov}
	\begin{equation}
		\mathrm{span}\left\{
		\boldsymbol{K}_{\sigma_1}^{-1} \! \boldsymbol{B} \, \boldsymbol{r}_1, \ldots, \boldsymbol{K}_{\sigma_r}^{-1} \! \boldsymbol{B} \, \boldsymbol{r}_r\right\} \subseteq \mathrm{ran}(\boldsymbol{V}),
	\end{equation}
	\begin{equation}
		\mathrm{span}\left\{
		\boldsymbol{K}_{\mu_1}^{- \! \mathsf T}\boldsymbol{C}^{\mathsf T} \boldsymbol{l}_1, \ldots, \boldsymbol{K}_{\mu_r}^{- \! \mathsf T}\boldsymbol{C}^{\mathsf T} \boldsymbol{l}_r\right\} \subseteq \mathrm{ran}(\boldsymbol{W}),
	\end{equation}
\end{subequations}
yielding the following tangential multipoint moment matching conditions:
\begin{equation}
\begin{aligned} \label{eq:tang_multipoint}
\boldsymbol{G}(\sigma_i) \, \boldsymbol{r}_i &= \boldsymbol{G}_{\mathrm{r}}(\sigma_i) \, \boldsymbol{r}_i, &i=1,\ldots,r, \\
\boldsymbol{l}_i^{\mathsf T} \boldsymbol{G}(\mu_i) &= \boldsymbol{l}_i^{\mathsf T} \, \boldsymbol{G}_{\mathrm{r}}(\sigma_i), &i=1,\ldots,r.
\end{aligned}
\end{equation}
Here, convenient right and left tangential directions $\! \boldsymbol{r}_i \!\in\! \mathbb{C}^m$ and $\boldsymbol{l}_i \in \mathbb{C}^p$ should be chosen. Besides, the shifts $\sigma_i$, $\mu_i \in \mathbb{C}$ cannot be quadratic eigenvalues of $(\boldsymbol{M}, \boldsymbol{D}, \boldsymbol{K})$.

\vspace{-0.2em}
\subsection{Equivalence of Krylov subspaces and Sylvester equations}
\vspace{-0.5em}
Any basis of an input and output Krylov subspace \eqref{eq:multipoint-tang-Krylov} can be interpreted as the solution $\boldsymbol{V}$ and $\boldsymbol{W}$ of the following second-order Sylvester equations:
\begin{subequations}
	\begin{align}
	\boldsymbol{M} \, \boldsymbol{V} \, \boldsymbol{S}_{v}^2 + \boldsymbol{D} \, \boldsymbol{V} \, \boldsymbol{S}_{v} + \boldsymbol{K} \, \boldsymbol{V} &= \boldsymbol{B} \, \boldsymbol{R}\,, \label{eq:Sylv-V} \\
	\boldsymbol{M}^{\mathsf{T}} \, \boldsymbol{W} \, \boldsymbol{S}_{w}^{2 \, \mathsf T} + \boldsymbol{D}^{\mathsf T} \, \boldsymbol{W} \, \boldsymbol{S}_{w}^{\mathsf T} + \boldsymbol{K}^{\mathsf{T}} \, \boldsymbol{W} &= \boldsymbol{C}^{\mathsf{T}} \, \boldsymbol{L}.
	\end{align}
\end{subequations}
Hereby, the input interpolation data $\left\{\sigma_i, \boldsymbol{r}_i\right\}$ is specified by the matrices $\boldsymbol{S}_v \!=\! \mathrm{diag}(\sigma_1, \ldots, \sigma_r) \!\in\! \mathbb{C}^{r \times r}$ and $\boldsymbol{R} \!=\! \left[\boldsymbol{r}_1, \ldots, \boldsymbol{r}_r\right] \in \mathbb{C}^{m \times r}$, where the pair $(\boldsymbol{R}, \boldsymbol{S}_v)$ is observable. Similarly, the output interpolation data $\left\{\mu_i, \boldsymbol{l}_i\right\}$ is denoted by the matrices $\boldsymbol{S}_w \!=\! \mathrm{diag}(\mu_1, \ldots, \mu_r) \in \mathbb{C}^{r \times r}$ and $\boldsymbol{L} \!=\! \left[\boldsymbol{l}_1, \ldots, \boldsymbol{l}_r \right] \in \mathbb{C}^{p \times r}$, where the pair $(\boldsymbol{S}_w, \boldsymbol{L}^{\mathsf T})$ is controllable.


\vspace{-0.2em}
\subsection{Time domain interpretation of Moment Matching}
\vspace{-0.5em}
In addition to the frequency domain perception of moment matching (cf. \eqref{eq:tang_multipoint}), one can also interpret this concept in the time domain. 
\begin{theorem} \label{th:lin-moments-steady-state}
	Consider the interconnection of system \eqref{eq:linear-FOM} with the linear signal generator
	\begin{equation} \label{eq:lin-SG}
	\begin{aligned}
	\dot{\boldsymbol{q}}_{\mathrm{r}}^v(t) &= \boldsymbol{S}_{v} \, \boldsymbol{q}_{\mathrm{r}}^v(t), \quad \boldsymbol{q}_{\mathrm{r}}^v(0) = \boldsymbol{q}_{\mathrm{r},0}^v \neq \boldsymbol{0}, \\
	\ddot{\boldsymbol{q}}_{\mathrm{r}}^v(t) &= \boldsymbol{S}_{v} \, \dot{\boldsymbol{q}}_{\mathrm{r}}^v(t), \quad \dot{\boldsymbol{q}}_{\mathrm{r}}^v(0) = \dot{\boldsymbol{q}}_{\mathrm{r},0}^v \neq \boldsymbol{0}, \\
	\boldsymbol{F}(t) &= \boldsymbol{R} \, \boldsymbol{q}_{\mathrm{r}}^v(t),
	\end{aligned}
	\end{equation}
	where the triple $(\boldsymbol{S}_v, \, \boldsymbol{R}, \, \boldsymbol{q}_{\mathrm{r},0}^v)$ is such that $(\boldsymbol{R}, \boldsymbol{S}_v)$ is observable, $\uplambda(\boldsymbol{S}_v) \, \cap \, \uplambda^2(\boldsymbol{M}, \boldsymbol{D}, \boldsymbol{K}) \!=\! \emptyset$ and $(\boldsymbol{S}_v, \boldsymbol{q}_{\mathrm{r},0}^v)$ is excitable. Let $\boldsymbol{V}$ be the solution of \eqref{eq:Sylv-V} and $\boldsymbol{W}$ such that $\det(\boldsymbol{W}^{\mathsf T} \boldsymbol{E} \boldsymbol{V}) \neq 0$. Furthermore, let $\boldsymbol{q}_{0} \!=\! \boldsymbol{V} \boldsymbol{q}_{\mathrm{r},0}^v$, $\dot{\boldsymbol{q}}_0 \!=\! \boldsymbol{V} \dot{\boldsymbol{q}}_{\mathrm{r},0}^v$ with $\boldsymbol{q}_{\mathrm{r},0}^v \!\neq\! \boldsymbol{0}$, $\dot{\boldsymbol{q}}_{\mathrm{r},0}^v \!\neq\! \boldsymbol{0}$ arbitrary. Then, the (asymptotically stable) ROM \eqref{eq:linear-ROM} exactly matches the steady-state response of the output of the FOM, i.e. $\boldsymbol{e}(t) \!=\! \boldsymbol{y}(t) - \boldsymbol{y}_{\mathrm{r}}(t) \!=\! \boldsymbol{C} \boldsymbol{q}(t) - \boldsymbol{C} \boldsymbol{V} \boldsymbol{q}_{\mathrm{r}}(t)\!=\! \boldsymbol{0} \ \forall \, t$ (see Fig. \ref{fig:lin-sys-lin-SG_steady-state-V}).
\end{theorem}
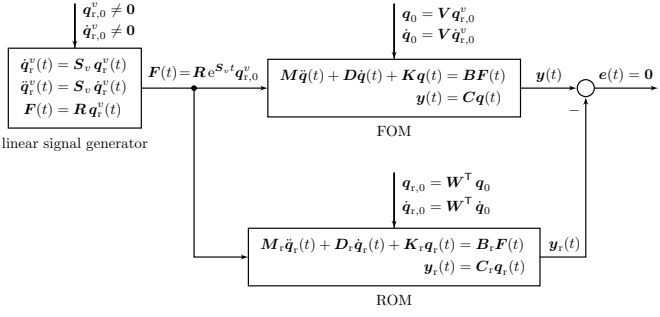
\begin{figure}[tp]
	\centering
	\scalebox{0.56}{
		\begin{tikzpicture}

  \UeFunk[align = left, inner sep = 6 pt]{linSigGen}{-0.8, 0}{0.6 cm}{
		$\begin{aligned}
		\dot{\boldsymbol{q}}_{\mathrm{r}}^v(t) &= \boldsymbol{S}_{v} \, \boldsymbol{q}_{\mathrm{r}}^v(t) \\[0.2em]
		\ddot{\boldsymbol{q}}_{\mathrm{r}}^v(t) &= \boldsymbol{S}_{v} \, \dot{\boldsymbol{q}}_{\mathrm{r}}^v(t) \\[0.2em]
		\boldsymbol{F}(t) &= \boldsymbol{R} \, \boldsymbol{q}_{\mathrm{r}}^v(t)
		\end{aligned}$
  }
  \UeFunk[align = left, inner sep = 6 pt]{FOM}{6.7, 0}{0.6 cm}{
	    $\begin{aligned}
	    \boldsymbol{M} \ddot{\boldsymbol{q}}(t) + \boldsymbol{D} \dot{\boldsymbol{q}}(t) + \boldsymbol{K} \boldsymbol{q}(t) &= \boldsymbol{B} \boldsymbol{F}(t) \\[0.2em]
	    \boldsymbol{y}(t) &= \boldsymbol{C} \boldsymbol{q}(t)
	    \end{aligned}$
  } 
  \UeFunk[align = left, inner sep = 6 pt]{ROM}{6.7, -4}{0.6 cm}{
	  	$\begin{aligned}
		\boldsymbol{M}_{\mathrm{r}} \ddot{\boldsymbol{q}}_{\mathrm{r}}(t) + \boldsymbol{D}_{\mathrm{r}} \dot{\boldsymbol{q}}_{\mathrm{r}}(t) + \boldsymbol{K}_{\mathrm{r}} \boldsymbol{q}_{\mathrm{r}}(t) &= \boldsymbol{B}_{\mathrm{r}} \boldsymbol{F}(t) \\[0.2em]
		\boldsymbol{y}_{\mathrm{r}}(t) &= \boldsymbol{C}_{\mathrm{r}} \boldsymbol{q}_{\mathrm{r}}(t)
     	\end{aligned}$
  }
  
  \Verzweigung{verzw-u}{2, 0}{3 pt}
  
  \Summationsstelle{summation}{11.2, 0}{0.4 cm}

  \draw[thick, -latex'] (linSigGen) -- (FOM) node[pos = 0.49, above] {$\boldsymbol{F}(t) \!=\! \boldsymbol{R} \, \mathrm{e}^{\boldsymbol{S}_v t} \boldsymbol{q}_{\mathrm{r},0}^v$};
  \draw[thick, -latex'] (verzw-u) |- (ROM);
  \draw[thick, -latex'] (FOM.east) -- (summation) node[pos = 0.51, above] {$\boldsymbol{y}(t)$};
  \draw[thick, -latex'] (ROM.east) -| (summation) node[pos = 0.27, above] {$\boldsymbol{y}_{\mathrm{r}}(t)$};
  \draw[thick, -latex'] (summation) -- + (1.7, 0) node[pos = 0.49, above] {$\boldsymbol{e}(t)=\boldsymbol{0}$};
  
  \draw[thick, -latex'] (linSigGen) -- + (0,2.0) node[pos = 0.6, right] {$\begin{aligned}
  	\boldsymbol{q}_{\mathrm{r},0}^v &\neq \boldsymbol{0} \\
	\dot{\boldsymbol{q}}_{\mathrm{r},0}^v &\neq \boldsymbol{0}
  	\end{aligned}$} -- (linSigGen);

  \draw[thick, -latex'] (FOM) -- + (0,2.0) node[pos = 0.6, right] {$\begin{aligned}
  	\boldsymbol{q}_{0} &= \boldsymbol{V} \boldsymbol{q}_{\mathrm{r},0}^v \\
  	\dot{\boldsymbol{q}}_{0} &= \boldsymbol{V} \dot{\boldsymbol{q}}_{\mathrm{r},0}^v
  	\end{aligned}$} -- (FOM);
  
  \draw[thick, -latex'] (ROM) -- + (0,2.0) node[pos = 0.6, right] {$\begin{aligned}
  	\boldsymbol{q}_{\mathrm{r},0} &= \boldsymbol{W}^{\mathsf T} \, \boldsymbol{q}_{0} \\
  	\dot{\boldsymbol{q}}_{\mathrm{r},0} &= \boldsymbol{W}^{\mathsf T} \, \dot{\boldsymbol{q}}_{0}
    \end{aligned}$} -- (ROM);

  \node[outer sep = 0.1 cm, below] at (linSigGen.south) {\text{linear signal generator}};
  \node[outer sep = 0.1 cm, below] at (FOM.south) {\text{FOM}};
  \node[outer sep = 0.1 cm, below] at (ROM.south) {\text{ROM}};
  
  \node[outer sep = 0.3 cm, below] at (summation.west) {$-\ $};
  
\end{tikzpicture}}
	\vspace{-1.7em}
	\caption{\footnotesize Interconnection between the linear FOM/ROM and the linear signal generator to illustrate the time domain interpretation of moment matching for linear structural systems.}
	\label{fig:lin-sys-lin-SG_steady-state-V}
\end{figure} 
\begin{corollary}
	Interconnecting system \eqref{eq:linear-FOM} with the signal ge\-nerator \eqref{eq:lin-SG} is equivalent to exciting the FOM with exponential input signals $\boldsymbol{F}(t) \!=\! \boldsymbol{R} \, \boldsymbol{q}_{\mathrm{r}}^v(t) \!=\! \boldsymbol{R} \, \mathrm{e}^{\boldsymbol{S}_v t} \, \boldsymbol{q}_{\mathrm{r},0}^v$ with exponents given by the shift matrix $\boldsymbol{S}_v$. Consequently, moment matching for linear structural systems can be interpreted as the \emph{exact} matching of the steady-state response of the FOM 
	\begin{equation}
	\begin{aligned}
	\boldsymbol{y}_{\textrm{SS}}(t) &= \boldsymbol{C} \sum_{i=1}^{r} \underbrace{(\sigma_i^2 \boldsymbol{M} + \sigma_i \boldsymbol{D} + \boldsymbol{K})^{-1} \boldsymbol{B} \boldsymbol{r}_i}_{\boldsymbol{v}_i} \underbrace{\mathrm{e}^{\sigma_i t} q_{\mathrm{r},0,i}^v}_{q_{\mathrm{r},i}^v(t)} \\
	&\equiv \boldsymbol{y}_{\mathrm{r}}(t) = \boldsymbol{C} \boldsymbol{V} \boldsymbol{q}_{\mathrm{r}}^v(t), \ \ \ \ \boldsymbol{m}_0(\sigma_i, \boldsymbol{r}_i) \!=\! \boldsymbol{C}\,\boldsymbol{v}_i,
	\end{aligned}
	\end{equation}
	when both FOM and ROM are excited with exponential input signals $\boldsymbol{F}(t) \!=\! \boldsymbol{R} \, \boldsymbol{q}_{\mathrm{r}}^v(t) \!=\! \boldsymbol{R} \, \mathrm{e}^{\boldsymbol{S}_v t} \, \boldsymbol{q}_{\mathrm{r},0}^v$ (see Fig. \ref{fig:lin-sys-lin-SG_steady-state-V}). For other input signals, the steady-state response is \emph{interpolated}. Note that the transient response of the FOM is also matched, if the initial conditions are chosen like above. 
\end{corollary}
The Sylvester equation \eqref{eq:Sylv-V} can be derived using the notion of signal generators. To this end, first insert the linear approximation ansatz $\boldsymbol{q}(t) \!=\! \boldsymbol{V} \boldsymbol{q}_{\mathrm{r}}(t)$ with $\boldsymbol{q}_{\mathrm{r}}(t) \!\overset{!}{=}\! \boldsymbol{q}_{\mathrm{r}}^v(t)$ in the state equation of~\eqref{eq:linear-FOM}. Then, the linear signal generator \eqref{eq:lin-SG} is plugged in, yielding
\begin{equation} \label{eq:derivation-Syl-2}
\left(\boldsymbol{M} \, \boldsymbol{V} \, \boldsymbol{S}_v^2 + \boldsymbol{D} \, \boldsymbol{V} \, \boldsymbol{S}_v + \boldsymbol{K} \, \boldsymbol{V} - \boldsymbol{B} \, \boldsymbol{R}\right) \cdot \boldsymbol{q}_{\mathrm{r}}^v(t) = \mathbf{0}.
\end{equation}
Since \eqref{eq:derivation-Syl-2} holds for $\boldsymbol{q}_{\mathrm{r}}^v(t) \!\!=\! \mathrm{e}^{\boldsymbol{S}_v t}\boldsymbol{q}_{\mathrm{r},0}^v$, the state vector $\boldsymbol{q}_{\mathrm{r}}^v(t)$ can be factored out and the \emph{constant} (state-independent) linear Sylvester equation \eqref{eq:Sylv-V} of dimension $n \!\times\! r$ is obtained.

\vspace{-0.6em}
\section{Moment Matching for nonlinear structural systems} \label{sec:nlmm-PDE}
\vspace{-0.6em}

Consider now a large-scale, nonlinear time-invariant, exponentially stable, MIMO second-order model of the form
\begin{align} \label{eq:nonlin-FOM}
\boldsymbol{M} \ddot{\boldsymbol{q}}(t) + \boldsymbol{D} \dot{\boldsymbol{q}}(t) + \boldsymbol{f}\big(\boldsymbol{q}(t)\big) &= \boldsymbol{B} \boldsymbol{F}(t), \ \boldsymbol{q}(0) \!=\! \boldsymbol{q}_0, \dot{\boldsymbol{q}}(0) \!=\! \dot{\boldsymbol{q}}_0, \nonumber \\
\boldsymbol{y}(t) &= \boldsymbol{C} \boldsymbol{q}(t),
\end{align}
with $\boldsymbol{q}(t) \in \mathbb{R}^n$, $\boldsymbol{F}(t) \in \mathbb{R}^m$, $\boldsymbol{y}(t) \in \mathbb{R}^p$ and the smooth mapping $\boldsymbol{f}(\boldsymbol{q})\!:\! \mathbb{R}^n \to \mathbb{R}^n$. 
Note that the modeling of damping in nonlinear dynamic analysis is not a trivial task. Thus, zero damping ($\boldsymbol{D} \!\!=\!\! \mathbf{0}$) or a linear Rayleigh damping $\boldsymbol{D} \!=\!\! \alpha \boldsymbol{M} + \beta \boldsymbol{K}(\boldsymbol{q}_0)$, $\boldsymbol{K}(\boldsymbol{q}_0) \!=\! \left. \frac{\partial \boldsymbol{f}(\boldsymbol{q})}{\partial \boldsymbol{q}} \right|_{\boldsymbol{q}=\boldsymbol{q}_0}$ are often assumed.  

\vspace{-0.7em}
\subsection{Nonlinear vs. Linear (Petrov-)Galerkin projection}
\vspace{-0.5em}
The goal is to find a nonlinear ROM of dimension $r \ll n$ using again a projection framework. 

One possibility consists in applying a nonlinear projection, where the approximation ansatz is given by $\boldsymbol{q}(t) \!=\! \boldsymbol{\nu}\big(\boldsymbol{q}_{\mathrm{r}}(t)\big)$ with the smooth nonlinear mapping $\boldsymbol{\nu}(\boldsymbol{q}_{\mathrm{r}}):\mathbb{R}^r \to \mathbb{R}^n$. Then,\vspace{-0.3em} 
\begin{equation*}
\dot{\boldsymbol{q}} = \frac{\partial \boldsymbol{\nu}(\boldsymbol{q}_{\mathrm{r}})}{\partial \boldsymbol{q}_{\mathrm{r}}} \, \dot{\boldsymbol{q}}_{\mathrm{r}}, \ \ \ddot{\boldsymbol{q}} = \frac{\partial \boldsymbol{\nu}(\boldsymbol{q}_{\mathrm{r}})}{\partial \boldsymbol{q}_{\mathrm{r}}} \, \ddot{\boldsymbol{q}}_{\mathrm{r}} + \frac{\partial^2 \boldsymbol{\nu}(\boldsymbol{q}_{\mathrm{r}})}{\partial \boldsymbol{q}_{\mathrm{r}}^2} \, \dot{\boldsymbol{q}}_{\mathrm{r}} \otimes \dot{\boldsymbol{q}}_{\mathrm{r}},
\end{equation*}
with the Jacobian $\widetilde{\boldsymbol{V}}(\boldsymbol{q}_{\mathrm{r}}) \!=\! \partial \boldsymbol{\nu}(\boldsymbol{q}_{\mathrm{r}})/\partial \boldsymbol{q}_{\mathrm{r}} \in \mathbb{R}^{n\times r}$ and the second derivative (Hessian) $\partial^2 \boldsymbol{\nu}(\boldsymbol{q}_{\mathrm{r}})/\partial \boldsymbol{q}_{\mathrm{r}}^2 \in \mathbb{R}^{n \times r^2}$. Inserting the ansatz and its derivatives in \eqref{eq:nonlin-FOM} yields an overdetermined systems of equations with the residual $\boldsymbol{\varepsilon}$. To obtain a quadratic system, we define $\boldsymbol{q}_{\mathrm{r}}(t) \!=\! \boldsymbol{\omega}\big(\boldsymbol{q}(t)\big)$ with smooth mapping $\boldsymbol{\omega}(\boldsymbol{q}):\mathbb{R}^n \to \mathbb{R}^r$ and premultiply the equations with the Jacobian $\widetilde{\boldsymbol{W}}(\boldsymbol{q})^{\mathsf T} \!=\! \left.\frac{\partial \boldsymbol{\omega}(\boldsymbol{q})}{\partial \boldsymbol{q}}\right|_{\boldsymbol{q}=\boldsymbol{\nu}(\boldsymbol{q}_{\mathrm{r}})}$, thus enforcing the Petrov-Galerkin condition $\widetilde{\boldsymbol{W}}(\boldsymbol{q})^{\mathsf T} \boldsymbol{\varepsilon} \!=\! \mathbf{0}$. This yields the ROM
\begin{equation} \label{eq:nonlin-ROM-nonlin-proj}
\begin{aligned}
\!\!\!\!\widetilde{\boldsymbol{M}}_{\mathrm{r}} \ddot{\boldsymbol{q}}_{\mathrm{r}} \, + \, \widetilde{\boldsymbol{g}} \, + \, \widetilde{\boldsymbol{D}}_{\mathrm{r}} \dot{\boldsymbol{q}}_{\mathrm{r}} \, + \, \widetilde{\boldsymbol{W}}(\boldsymbol{q})^{\mathsf T} \boldsymbol{f}\big(\boldsymbol{\nu}(\boldsymbol{q}_{\mathrm{r}})\big) &= \widetilde{\boldsymbol{B}}_{\mathrm{r}} \boldsymbol{F},\\
\boldsymbol{y}_{\mathrm{r}} &= \boldsymbol{C} \, \boldsymbol{\nu}\big(\boldsymbol{q}_{\mathrm{r}}\big),
\end{aligned}
\end{equation}
with $\left\{\widetilde{\boldsymbol{M}}_{\mathrm{r}}, \widetilde{\boldsymbol{D}}_{\mathrm{r}}\right\} \!=\! \widetilde{\boldsymbol{W}}(\boldsymbol{q})^{\mathsf T} \!\left\{\boldsymbol{M}, \boldsymbol{D}\right\} \widetilde{\boldsymbol{V}}(\boldsymbol{q}_{\mathrm{r}})$, $\widetilde{\boldsymbol{B}}_{\mathrm{r}} \!=\! \widetilde{\boldsymbol{W}}(\boldsymbol{q})^{\mathsf T} \boldsymbol{B}$, the convective term $\widetilde{\boldsymbol{g}} \!=\! \widetilde{\boldsymbol{W}}(\boldsymbol{q})^{\mathsf T} \boldsymbol{M} \frac{\partial^2 \boldsymbol{\nu}(\boldsymbol{q}_{\mathrm{r}})}{\partial \boldsymbol{q}_{\mathrm{r}}^2} \, \dot{\boldsymbol{q}}_{\mathrm{r}} \otimes \dot{\boldsymbol{q}}_{\mathrm{r}}$ and the initial conditions $\boldsymbol{q}_{\mathrm{r}}(0) \!=\! \widetilde{\boldsymbol{W}}(\boldsymbol{q})^{\mathsf T} \, \boldsymbol{q}_0, \, \dot{\boldsymbol{q}}_{\mathrm{r}}(0) \!=\! \widetilde{\boldsymbol{W}}(\boldsymbol{q})^{\mathsf T} \, \dot{\boldsymbol{q}}_0$.

Another possibility to reduce the nonlinear system \eqref{eq:nonlin-FOM} is to apply the common and well-known linear Petrov-Galerkin projection, where linear mappings given by the matrices $\boldsymbol{V}, \boldsymbol{W}$ are used instead. This yields the ROM 
\begin{equation} \label{eq:nonlin-ROM}
\begin{aligned}
\boldsymbol{M}_{\mathrm{r}} \ddot{\boldsymbol{q}}_{\mathrm{r}}(t) + \boldsymbol{D}_{\mathrm{r}} \dot{\boldsymbol{q}}_{\mathrm{r}}(t) + \boldsymbol{W}^{\mathsf{T}} \boldsymbol{f}\big(\boldsymbol{V} \boldsymbol{q}_{\mathrm{r}}(t)\big) &= \boldsymbol{B}_{\mathrm{r}} \boldsymbol{F}(t), \\
\boldsymbol{y}_{\mathrm{r}}(t) &= \boldsymbol{C}_{\mathrm{r}} \boldsymbol{q}_{\mathrm{r}}(t),
\end{aligned}
\end{equation}
with reduced matrices and initial conditions given as in the linear case.
\begin{remark}\label{rem:LPvsNLP}
	Please note that the use of a linear projection $\boldsymbol{q} \!\approx\! \boldsymbol{V} \boldsymbol{q}_{\mathrm{r}}$ constitutes a special case of the most general nonlinear projection $\boldsymbol{q} \!\approx\! \boldsymbol{\nu}(\boldsymbol{q}_{\mathrm{r}})$ or the power series ansatz 
	\begin{equation}\label{eq:power_series_projection}
	\boldsymbol{q} = \sum_{k=1}^{N} \boldsymbol{V}^{(k)} \, \boldsymbol{q}_{\mathrm{r}}^{(k)} = \boldsymbol{V}^{(1)} \boldsymbol{q}_{\mathrm{r}}^{(1)} + \boldsymbol{V}^{(2)} \boldsymbol{q}_{\mathrm{r}}^{(2)} + \cdots 	\vspace{-1.0em}
	\end{equation}
	with $\boldsymbol{V}^{(k)} \in \mathbb{R}^{n \times r^k}$ and $\boldsymbol{q}_{\mathrm{r}}^{(k)} \!=\! \overbrace{\boldsymbol{q}_{\mathrm{r}} \otimes \cdots \otimes \boldsymbol{q}_{\mathrm{r}}}^{k \, \mathrm{times}} \in \mathbb{R}^{r^k}$. 
\end{remark}

\vspace{-0.5em}
\subsection{Steady-State-Based Nonlinear Moment Matching}
\vspace{-0.5em}
The steady-state-based interpretation of moment matching can be carried over to nonlinear structural systems. 
\begin{theorem} \label{th:nonlin-moments-steady-state}
	Consider the interconnection of system \eqref{eq:nonlin-FOM} with the nonlinear signal generator
	\begin{align} \label{eq:nonlin-SG}
	\dot{\boldsymbol{q}}_{\mathrm{r}}^v(t) &= \boldsymbol{s}_{v}\big(\boldsymbol{q}_{\mathrm{r}}^v(t)\big), \quad \boldsymbol{q}_{\mathrm{r}}^v(0) = \boldsymbol{q}_{\mathrm{r},0}^v \neq \boldsymbol{0}, \nonumber \\[0.2em]
	\ddot{\boldsymbol{q}}_{\mathrm{r}}^v(t) &= \frac{\partial \boldsymbol{s}_v\big(\boldsymbol{q}_{\mathrm{r}}^v(t)\big)}{\partial \boldsymbol{q}_{\mathrm{r}}^v(t)} \cdot \boldsymbol{s}_{v}\big(\boldsymbol{q}_{\mathrm{r}}^v(t)\big), \quad \dot{\boldsymbol{q}}_{\mathrm{r}}^v(0) = \dot{\boldsymbol{q}}_{\mathrm{r},0}^v \neq \boldsymbol{0}, \nonumber \\[0.2em]
	\boldsymbol{F}(t) &= \boldsymbol{r}\big(\boldsymbol{q}_{\mathrm{r}}^v(t)\big),
	\end{align}
	where $\boldsymbol{s}_{v}(\boldsymbol{q}_{\mathrm{r}}^v) \!:\! \mathbb{R}^r \to \mathbb{R}^r$, $\boldsymbol{r}(\boldsymbol{q}_{\mathrm{r}}^v) \!:\! \mathbb{R}^r \to \mathbb{R}^m$ are smooth mappings. Let $\boldsymbol{\nu}(\boldsymbol{q}_{\mathrm{r}}^v)$ be the unique solution of the following Sylvester-like partial differential equation~(PDE)
	\begin{align} \label{eq:nonlin-time-dep-Sylv-v}
	\!\!\boldsymbol{M} & \frac{\partial \boldsymbol{\nu}(\boldsymbol{q}_{\mathrm{r}}^v)}{\partial \boldsymbol{q}_{\mathrm{r}}^v} \, \frac{\partial \boldsymbol{s}_v\big(\boldsymbol{q}_{\mathrm{r}}^v\big)}{\partial \boldsymbol{q}_{\mathrm{r}}^v} \, \boldsymbol{s}_v(\boldsymbol{q}_{\mathrm{r}}^v) + \boldsymbol{M} \frac{\partial^2 \boldsymbol{\nu}(\boldsymbol{q}_{\mathrm{r}}^v)}{\partial \boldsymbol{q}_{\mathrm{r}}^{v 2}} \, \boldsymbol{s}_v(\boldsymbol{q}_{\mathrm{r}}^v) \otimes \boldsymbol{s}_v(\boldsymbol{q}_{\mathrm{r}}^v) \nonumber \\[0.2em]
	&+ \boldsymbol{D} \frac{\partial \boldsymbol{\nu}(\boldsymbol{q}_{\mathrm{r}}^v)}{\partial \boldsymbol{q}_{\mathrm{r}}^v} \, \boldsymbol{s}_v(\boldsymbol{q}_{\mathrm{r}}^v) + \boldsymbol{f}\big(\boldsymbol{\nu}(\boldsymbol{q}_{\mathrm{r}}^v)\big) = \boldsymbol{B} \, \boldsymbol{r}\big(\boldsymbol{q}_{\mathrm{r}}^v\big).
	\end{align}
	and $\boldsymbol{\omega}(\boldsymbol{q})$ such that $\boldsymbol{\omega}(\boldsymbol{\nu}(\boldsymbol{q}_{\mathrm{r}}^v)) \!=\! \boldsymbol{q}_{\mathrm{r}}^v$. Furthermore, let $\boldsymbol{q}_0 \!=\! \boldsymbol{\nu}(\boldsymbol{q}_{\mathrm{r},0}^v)$, $\dot{\boldsymbol{q}}_0 \!=\! \boldsymbol{\nu}(\dot{\boldsymbol{q}}_{\mathrm{r},0}^v)$ with $\boldsymbol{q}_{\mathrm{r},0}^v \!\neq\! \boldsymbol{0}$, $\dot{\boldsymbol{q}}_{\mathrm{r},0}^v \!\neq\! \boldsymbol{0}$ arbitrary. Then, the (exponentially stable) ROM \eqref{eq:nonlin-ROM-nonlin-proj} exactly matches the steady-state response of the output of the FOM, i.e. $\boldsymbol{e}(t) \!=\! \boldsymbol{y}(t) - \boldsymbol{y}_{\mathrm{r}}(t) \!=\! \boldsymbol{C} \boldsymbol{q}(t) - \boldsymbol{C}\boldsymbol{\nu}\big(\boldsymbol{q}_{\mathrm{r}}(t)\big) \!=\! \boldsymbol{0} \ \forall \, t$.
\end{theorem}
\begin{corollary}	
	The $0$-th moment of system \eqref{eq:nonlin-FOM} at $(\boldsymbol{s}_v(\boldsymbol{q}_{\mathrm{r}}^v),$ $\boldsymbol{r}(\boldsymbol{q}_{\mathrm{r}}^v), \boldsymbol{q}_{\mathrm{r},0}^v)$ is related to the (locally well-defined) steady-state response of the FOM $\boldsymbol{y}_{\textrm{SS}}(t) \!=\! \boldsymbol{C}\boldsymbol{\nu}\big(\boldsymbol{q}_{\mathrm{r}}^v(t)\big)$. Thus, nonlinear moment matching can be interpreted as the \emph{exact} matching of the steady-state response of the FOM
	\begin{equation}
	\begin{aligned}
	\!\!\!\boldsymbol{y}_{\textrm{SS}}(t) = \boldsymbol{C} \, \boldsymbol{q}_{\textrm{SS}}(t) &\equiv \boldsymbol{y}_{\mathrm{r}}(t) = \boldsymbol{C} \boldsymbol{\nu}\big(\boldsymbol{q}_{\mathrm{r}}^v(t)\big) \\
	&:= \boldsymbol{m}_0\big(\boldsymbol{s}_v(\boldsymbol{q}_{\mathrm{r}}^v(t)), \boldsymbol{r}(\boldsymbol{q}_{\mathrm{r}}^v(t)), \boldsymbol{q}_{\mathrm{r},0}^v\big),
	\end{aligned}
	\end{equation}
	when both FOM and ROM are excited with the signal generator \eqref{eq:nonlin-SG}. For other input signals, the steady-state response is \emph{interpolated}. Please note here again that the transient response of the FOM is also matched, if the initial conditions are chosen like above. In such case, the matching conditions hold for all $t$ (transient+steady-state). Otherwise, they hold only for $t \to \infty$ (steady-state).
\end{corollary}
The Sylvester-like PDE \eqref{eq:nonlin-time-dep-Sylv-v} represents the nonlinear counterpart of the linear equation \eqref{eq:derivation-Syl-2}. Thus, the PDE \eqref{eq:nonlin-time-dep-Sylv-v} has been derived similarly as follows. First, the nonlinear approximation ansatz $\boldsymbol{q}(t) \!=\! \boldsymbol{\nu}(\boldsymbol{q}_{\mathrm{r}}(t))$ with $\boldsymbol{q}_{\mathrm{r}}(t) \overset{!}{=} \boldsymbol{q}_{\mathrm{r}}^v(t)$ is inserted in the state equation of \eqref{eq:nonlin-FOM}. Afterwards, the nonlinear signal generator \eqref{eq:nonlin-SG} is plugged in, yielding \eqref{eq:nonlin-time-dep-Sylv-v}. 
Note that -- as opposed to the linear, state-independent Sylvester equation \eqref{eq:Sylv-V} of dimension $n \!\times\! r$ -- the PDE \eqref{eq:nonlin-time-dep-Sylv-v} is a \emph{nonlinear}, \emph{state-dependent} equation of dimension~$n \!\times\! 1$.

\vspace{-0.5em}
\section{Simulation-Free Reduction of Nonlinear Structural Systems} \label{sec:simulation-free-NLMM}
\vspace{-0.5em}
\subsection{Simplifications}
\vspace{-0.5em}
The approach for nonlinear moment matching described in Section \ref{sec:nlmm-PDE} requires the solution $\boldsymbol{\nu}(\boldsymbol{q}_{\mathrm{r}}^v(t))$ of the nonlinear, state-dependent PDE \eqref{eq:nonlin-time-dep-Sylv-v} to reduce the FOM \eqref{eq:nonlin-FOM}. We propose similar step-by-step simplifications as in \cite{cruz2019practicable} to achieve a practicable, simulation-free reduction method for nonlinear structural systems, which relies on solving nonlinear systems of equations rather than on the difficult solution of the nonlinear PDE \eqref{eq:nonlin-time-dep-Sylv-v}. The proposed simplifications are \textit{(i)} the use of a linear projection, \textit{(ii)} the column-wise consideration of the equation and \textit{(iii)} a time discretization with time-snapshots. In the following, we explain the simplifications for three different signal generator cases:




\vspace{-0.4em}
\subsection*{1.) Nonlinear signal generator}
\vspace{-0.5em}

\subsubsection*{(i) Linear projection}
Motivated by the fact that nonlinear projections are complicated, whereas linear ones are often successfully employed even in nonlinear MOR, we propose to apply a linear projection $\boldsymbol{q}(t) \!=\! \boldsymbol{\nu}(\boldsymbol{q}_{\mathrm{r}}^v(t)) \!=\! \boldsymbol{V} \, \boldsymbol{q}_{\mathrm{r}}^v(t)$ instead of the nonlinear projection mapping $\boldsymbol{\nu}(\boldsymbol{q}_{\mathrm{r}}^v(t))$. 


By doing so, the PDE \eqref{eq:nonlin-time-dep-Sylv-v} becomes the following \emph{algebraic} nonlinear system of equations
\begin{equation} \label{eq:LP-NSG}
\begin{aligned}
\boldsymbol{M} \boldsymbol{V} \, \frac{\partial \boldsymbol{s}_v\big(\boldsymbol{q}_{\mathrm{r}}^v(t)\big)}{\partial \boldsymbol{q}_{\mathrm{r}}^v(t)} &\, \boldsymbol{s}_v\big(\boldsymbol{q}_{\mathrm{r}}^v(t)\big) + \boldsymbol{D} \,  \boldsymbol{V} \, \boldsymbol{s}_v\big(\boldsymbol{q}_{\mathrm{r}}^v(t)\big) \\ 
&+ \boldsymbol{f}\big(\boldsymbol{V} \boldsymbol{q}_{\mathrm{r}}^v(t)\big) - \boldsymbol{B} \, \boldsymbol{r}\big(\boldsymbol{q}_{\mathrm{r}}^v(t)\big) = \mathbf{0},
\end{aligned}
\end{equation}
where the triple $\big(\boldsymbol{s}_v(\boldsymbol{q}_{\mathrm{r}}^v(t)), \boldsymbol{r}(\boldsymbol{q}_{\mathrm{r}}^v(t)), \boldsymbol{q}_{\mathrm{r}}^v(t)\big)$ is user-defined.

\vspace{-0.5em}
\subsubsection*{(ii) Column-wise consideration}
System \eqref{eq:LP-NSG} consists of $n$ equations for $n \cdot r$ unknowns in $\boldsymbol{V} \in \mathbb{R}^{n \times r}$, i.e. it is underdetermined. To overcome this problem, we propose to consider the equation column-wise for each $\boldsymbol{v}_i \in \mathbb{R}^n$, $i=1,\ldots,r$
\begin{equation} \label{eq:LP-NSG-elem}
\begin{aligned}
\boldsymbol{M} \, \boldsymbol{v}_i  \frac{\partial s_{v_{i}}\big(q_{\mathrm{r},i}^v(t)\big)}{\partial q_{\mathrm{r},i}^v(t)} &\, s_{v_{i}}\big(q_{\mathrm{r},i}^v(t)\big) + \boldsymbol{D} \, \boldsymbol{v}_i \,  s_{v_{i}}\big(q_{\mathrm{r},i}^v(t)\big) \\
&\!\!\!\!\!\!\!\!\!\!+ \boldsymbol{f}\big(\boldsymbol{v}_i \, q_{\mathrm{r}, i}^v(t)\big) - \boldsymbol{B} \, \boldsymbol{r}_{_i}\big(q_{\mathrm{r},i}^v(t)\big) = \mathbf{0},
\end{aligned}
\end{equation} 
with $q_{\mathrm{r},i}^v(t) \in \mathbb{R}$ and $\boldsymbol{V} \!=\! \left[\boldsymbol{v}_1, \ldots, \boldsymbol{v}_{r}\right]$. Please be aware that, in contrast to the linear setting, a column-wise construction of $\boldsymbol{V}$ using columns $\boldsymbol{v}_i$ satisfying \eqref{eq:LP-NSG-elem} does generally not fulfill the ``true" equation \eqref{eq:LP-NSG}.


\vspace{-0.5em}
\subsubsection*{(iii) Time discretization}
Different from the linear case (see \eqref{eq:derivation-Syl-2}), in the non\-li\-near setting $\boldsymbol{q}_{\mathrm{r}}^v(t)$ can generally not be factored out in \eqref{eq:nonlin-time-dep-Sylv-v} anymore. Consequently, the nonlinear equation \eqref{eq:LP-NSG-elem} is still state-dependent. Thus, inspired by POD, we propose to discretize the state-dependent equation with \emph{time-snapshots} $\left\{t^*_k\right\}$, $k=1,\ldots,K$.

With user-defined $s_{v_{i}}(q_{\mathrm{r},i}^v(t^*_k))$, $\boldsymbol{r}_{_i}(q_{\mathrm{r},i}^v(t^*_k))$ and $q_{\mathrm{r},0,i}^v$, the following time-independent equation results 
\begin{align} \label{eq:LP-NSG-elem-timeDis}
\boldsymbol{M} \, \boldsymbol{v}_{ik}  \frac{\partial s_{v_{i}}\big(q_{\mathrm{r},i}^v(t^*_k)\big)}{\partial q_{\mathrm{r},i}^v(t^*_k)} &\, s_{v_{i}}\big(q_{\mathrm{r},i}^v(t^*_k)\big) + \boldsymbol{D} \, \boldsymbol{v}_{ik} \,  s_{v_{i}}\big(q_{\mathrm{r},i}^v(t^*_k)\big) \nonumber \\
&\!\!\!\!\!\!\!\!\!\!\!\!\!\!\!\!\!\!\!\!\!\!\!\!+ \boldsymbol{f}\big(\boldsymbol{v}_{ik} \, q_{\mathrm{r}, i}^v(t^*_k)\big) - \boldsymbol{B} \, \boldsymbol{r}_{_i}\big(q_{\mathrm{r},i}^v(t^*_k)\big) = \mathbf{0},
\end{align}
which can be solved for each $\boldsymbol{v}_{ik} \in \mathbb{R}^n$, with $i=1,\ldots,r$ and $k=1,\ldots,K$. The discrete solution $q_{\mathrm{r},i}^v(t^*_k)$ of the nonlinear signal generator equation \eqref{eq:nonlin-SG} must be given or computed via simulation before solving equation \eqref{eq:LP-NSG-elem-timeDis}.

\vspace{-0.5em}
\subsection*{2.) Linear signal generator}
\vspace{-0.5em}
One may also come to the idea of interconnecting the nonlinear system \eqref{eq:nonlin-FOM} with the linear signal generator \eqref{eq:lin-SG}, where $\boldsymbol{s}_v(\boldsymbol{q}_{\mathrm{r}}^v(t)) \!=\! \boldsymbol{S}_v \, \boldsymbol{q}_{\mathrm{r}}^v(t)$ and $\boldsymbol{r}(\boldsymbol{q}_{\mathrm{r}}^v(t)) \!=\! \boldsymbol{R} \, \boldsymbol{q}_{\mathrm{r}}^v(t)$.
\vspace{-0.5em}
\subsubsection*{(i) Linear projection}
By doing so, equation \eqref{eq:LP-NSG} becomes
\begin{equation}\label{eq:LP-LSG}
\boldsymbol{M} \boldsymbol{V} \boldsymbol{S}_v^2 \boldsymbol{q}_{\mathrm{r}}^v(t) + \boldsymbol{D} \boldsymbol{V} \boldsymbol{S}_v \boldsymbol{q}_{\mathrm{r}}^v(t) + \boldsymbol{f}\big(\boldsymbol{V} \boldsymbol{q}_{\mathrm{r}}^v(t)\big) - \boldsymbol{B} \boldsymbol{R} \boldsymbol{q}_{\mathrm{r}}^v(t) = \mathbf{0}.
\end{equation}
where the triple $(\boldsymbol{S}_v, \, \boldsymbol{R}, \, \boldsymbol{q}_{\mathrm{r},0}^v)$ is user-defined. \\
Note that a linear signal generator corresponds to exciting the nonlinear system with exponential input signals~$\boldsymbol{F}(t) \!=\! \boldsymbol{R} \, \boldsymbol{q}_{\mathrm{r}}^v(t) \!=\! \boldsymbol{R} \, \mathrm{e}^{\boldsymbol{S}_v t} \, \boldsymbol{q}_{\mathrm{r},0}^v$. Although exponential functions are the characterizing eigenfunctions for linear systems, the question raises whether complex (growing) exponentials are sufficiently valid for characterizing nonlinear systems. Nevertheless, one might hope that such exponential input signals can be sufficient in certain cases to describe the underlying nonlinear dynamics adequately as well.

\vspace{-0.5em}   
\subsubsection*{(ii) Column-wise consideration}
Considering the underdetermined equation \eqref{eq:LP-LSG} again column-wise delivers
\begin{equation} \label{eq:LP-LSG-elem}
\begin{aligned}
\!\!\boldsymbol{M} \boldsymbol{v}_i \sigma_i^2 q_{\mathrm{r},i}^v(t) \!+\! \boldsymbol{D} \boldsymbol{v}_i \sigma_i q_{\mathrm{r},i}^v(t) \!+\! \boldsymbol{f}\big(\boldsymbol{v}_i q_{\mathrm{r}, i}^v(t)\big) \!-\! \boldsymbol{B}\boldsymbol{r}_i q_{\mathrm{r},i}^v(t) = \mathbf{0},
\end{aligned}
\end{equation}
where the signal generator \eqref{eq:lin-SG} becomes $\dot{q}_{\mathrm{r},i}^v(t) \!=\! \sigma_i \, q_{\mathrm{r},i}^v(t)$, $\boldsymbol{F}_i(t) \!=\! \boldsymbol{r}_i \, q_{\mathrm{r},i}^v(t)$ with $q_{\mathrm{r},i}^v(t) \!=\! \mathrm{e}^{\sigma_i t} q_{\mathrm{r},0,i}^v$ for $i=1,\ldots,r$.

\vspace{-0.5em}
\subsubsection*{(iii) Time discretization}
Using the time-discretized signal generator $\dot{q}_{\mathrm{r},i}^v(t^*_k) \!=\! \sigma_i \, q_{\mathrm{r},i}^v(t^*_k)$, $\boldsymbol{F}_i(t^*_k) \!=\! \boldsymbol{r}_i \, q_{\mathrm{r},i}^v(t^*_k)$ and $q_{\mathrm{r},0,i}^v$, equation \eqref{eq:LP-LSG-elem} becomes time-independent
\begin{equation} \label{eq:LP-LSG-elem-timeDis}
\begin{aligned}
\boldsymbol{M} \, \boldsymbol{v}_{ik}  &\, \sigma_i^2 \, q_{\mathrm{r},i}^v(t^*_k) + \boldsymbol{D} \, \boldsymbol{v}_{ik} \, \sigma_i \, q_{\mathrm{r},i}^v(t^*_k) \\
&+ \boldsymbol{f}\big(\boldsymbol{v}_{ik} \, q_{\mathrm{r}, i}^v(t^*_k)\big) - \boldsymbol{B} \, \boldsymbol{r}_i \, q_{\mathrm{r},i}^v(t^*_k) = \mathbf{0},
\end{aligned}
\end{equation}
with $q_{\mathrm{r},i}^v(t^*_k) \!=\! \mathrm{e}^{\sigma_i t^*_k} \, q_{\mathrm{r},0,i}^v$ for $i=1,\ldots,r$. Note that in this case, the discrete solution $q_{\mathrm{r},i}^v(t^*_k)$ of the linear signal generator equation \eqref{eq:lin-SG} is analytically given by exponential functions with exponents $\sigma_i$, so that no simulation of the linear signal generator is required.

\vspace{-0.5em}
\subsection*{3.) Zero signal generator}
\vspace{-0.5em}
This special (linear) signal generator is defined as $\dot{\boldsymbol{q}}_{\mathrm{r}}^v(t) \!=\! \boldsymbol{s}_v(\boldsymbol{q}_{\mathrm{r}}^v(t)) \!=\! \boldsymbol{0}$, yielding $\boldsymbol{q}_{\mathrm{r}}^v(t) \!=\! \boldsymbol{q}_{\mathrm{r},0}^v \!=\! \textrm{const}$ and $\boldsymbol{F}(t) \!=\! \boldsymbol{R} \boldsymbol{q}_{\mathrm{r}}^v(t) \!=\! \boldsymbol{R} \, \boldsymbol{q}_{\mathrm{r},0}^v \!=\! \textrm{const}$. Thus, this generator corresponds to exciting the nonlinear system with a constant input.

\vspace{-0.5em}
\subsubsection*{(i) Linear projection}
In this particular case, equation \eqref{eq:LP-NSG} becomes
\begin{equation} \label{eq:LP-ZSG}
\boldsymbol{f}\big(\boldsymbol{V} \boldsymbol{q}_{\mathrm{r},0}^v\big) - \boldsymbol{B} \, \boldsymbol{R} \, \boldsymbol{q}_{\mathrm{r},0}^v = \mathbf{0},
\end{equation}
i.e. a nonlinear, \emph{time-independent} system of equations.

\vspace{-0.5em}
\subsubsection*{(ii) Column-wise consideration}
A column-wise consideration of the above underdetermined equation yields 
\begin{equation} \label{eq:LP-ZSG-elem}
\boldsymbol{f}\big(\boldsymbol{v}_{i} \, q_{\mathrm{r},0,i}^v\big) - \boldsymbol{B} \, \boldsymbol{r}_i \, q_{\mathrm{r},0,i}^v = \mathbf{0},
\end{equation}
where $\dot{q}_{\mathrm{r},i}^v(t) \!=\! 0$ with $\sigma_i \!=\! 0$, $\boldsymbol{F}_i(t) \!=\! \boldsymbol{r}_i \, q_{\mathrm{r},0,i}^v \!=\! \textrm{const}$ and $q_{\mathrm{r},i}^v(t) \!=\! q_{\mathrm{r},0,i}^v \!=\! \textrm{const}$ hold for $i=1,\ldots,r$. In other words, the employment of a zero signal generator is equivalent to moment matching at shifts $\sigma_i \!=\! 0$. 

\vspace{-0.5em}
\subsubsection*{(iii) Time discretization}
For this special case, no time discretization is needed, since \eqref{eq:LP-ZSG-elem} already represents a time-independent equation. Please note that solving \eqref{eq:LP-ZSG-elem} is strong related to computing the \emph{steady-state} $\boldsymbol{q}_{\infty}$ (aka. \emph{equilibrium} or \emph{static displacement}) of the nonlinear mechanical system \eqref{eq:nonlin-FOM} by means of $\boldsymbol{f}\big(\boldsymbol{q}_{\infty}\big) \!-\! \boldsymbol{B}\boldsymbol{F}_{\mathrm{const}} \!=\! \mathbf{0}$.

\vspace{-0.5em}
\subsection{Simulation-free nonlinear moment matching algorithm}
\vspace{-0.5em}
Based on the simplifications discussed before, we now state our proposed simulation-free nonlinear moment matching algorithm for nonlinear structural systems:  
\begin{algorithm}[!ht]\caption{Second-order NLMM (SO-NLMM)} \label{alg:nlmm}
	\begin{algorithmic}[1]
		\Require $\boldsymbol{M}$, $\boldsymbol{D}$, $\boldsymbol{f}(\boldsymbol{q})$, $\boldsymbol{B}$, $\!\boldsymbol{J}_{\boldsymbol{f}}(\boldsymbol{q})$, $\!q_{\mathrm{r},i}^v(t^*_k)$, $\!\dot{q}_{\mathrm{r},i}^v(t^*_k)$, $\!\ddot{q}_{\mathrm{r},i}^v(t^*_k)$ $\!\boldsymbol{r}_{_i}(q_{\mathrm{r},i}^v(t^*_k))$, initial guesses $\boldsymbol{v}_{0,ik}$, deflated order $r_{\mathrm{defl}}$
		\Ensure orthogonal basis $\boldsymbol{V}$ \vspace{0.2em}
		\For{\begin{small} \texttt{i = 1 : r} \end{small}} \hspace{1.5em} $\triangleright$ e.g. $r$ different shifts $\sigma_i$
		\For{\begin{small} \texttt{k = 1 : K} \end{small}} \vspace{0.2em} \hspace{0.4em} $\triangleright$ e.g. $K$ samples in each shift 
		\State \begin{footnotesize} \hspace{-1.5em}
		\texttt{fun=@(v)}$\boldsymbol{M}\,\texttt{v}\,\ddot{q}_{\mathrm{r},i}^v(t^*_k) + \boldsymbol{D} \, \texttt{v} \, \dot{q}_{\mathrm{r},i}^v(t^*_k) + \boldsymbol{f}(\texttt{v}\,q_{\mathrm{r},ik}^v)  -\! \boldsymbol{B}\,\boldsymbol{r}_{_i}(q_{\mathrm{r},ik}^v)$ \end{footnotesize} \label{al:line:fun} \vspace{-0.5em}
		\State \begin{small} \hspace{-1.8em} \texttt{Jfun=@(v)} $\boldsymbol{M}\,\ddot{q}_{\mathrm{r},i}^v(t^*_k) + \boldsymbol{D}\,\dot{q}_{\mathrm{r},i}^v(t^*_k) + \boldsymbol{J}_{\boldsymbol{f}}(\texttt{v}\,q_{\mathrm{r},ik}^v)\,q_{\mathrm{r},ik}^v$ \end{small} \label{al:line:Jfun} \vspace{0.4em}		
		\State \begin{small} \hspace{-1.7em} \texttt{V(:,(i-1)*K+k)=} \textbf{\texttt{Newton}}\texttt{(fun,}$\, \boldsymbol{v}_{0,ik} \, $\texttt{,Jfun)} \end{small} \label{al:line:Newton} \vspace{0.3em}
		\State \begin{small} \hspace{-1.7em} \texttt{V = }\textbf{\texttt{gramSchmidt}}\texttt{((i-1)*K+k, V)} \end{small} \label{al:line:gramSchmidt} \vspace{0.2em} \hspace{0.1em} $\triangleright$ optional  
		\EndFor
		\EndFor
		\State \texttt{V = }\textbf{\texttt{svd}}\texttt{(V,}$\, r_{\mathrm{defl}}$\texttt{)} \hspace{1.1em} $\triangleright$ deflation is optional \label{al:line:SVD}
	\end{algorithmic}
\end{algorithm}\\
Since the different strategies, the computational effort and further aspects of the algorithm have been already discussed in detail for nonlinear first-order systems in \cite{cruz2019practicable}, here we briefly summarize these aspects for the second-order case, and rather provide some new clarifying insights that also hold true for the first-order case.


\vspace{-0.5em}
\subsection*{a) Different strategies, degrees of freedom and special cases}
\begin{itemize}[leftmargin=*]
	\item Besides the depicted most general approach using different signal generators and time-snapshots ($i\!=\!1,\ldots,r$, $k \!=\! 1, \ldots,K$), one could also consider a single signal generator at several collocation points ($i\!=\!1$, $k \!=\! 1, \ldots,K$) or different signal generators at only one time-snapshot ($i \!=\! 1,\ldots,r$, $K \!=\! 1$).
	\item In a\-ddi\-tion to a nonlinear signal generator, one could also apply a linear or a zero signal generator. To this end, line \ref{al:line:fun} (and correspondingly line \ref{al:line:Jfun} also) in Algorithm \ref{alg:nlmm} should be replaced by the equations \eqref{eq:LP-LSG-elem-timeDis} and \eqref{eq:LP-ZSG-elem}. 
	\item If the NLMM algorithm is applied to a \emph{linear} second-order system \eqref{eq:linear-ROM} with $\boldsymbol{f}(\boldsymbol{q}) \!=\! \boldsymbol{K} \boldsymbol{q}$ using a linear signal generator, the algorithm boils down to the classical rational Krylov subspace method, since line \ref{al:line:fun} becomes
	\begin{align}
	(\boldsymbol{M} \boldsymbol{v}_{ik} \, \sigma_i^2 &+ \boldsymbol{D} \boldsymbol{v}_{ik} \, \sigma_i + \boldsymbol{K}\boldsymbol{v}_{ik}) \, \cancel{\mathrm{e}^{\sigma_i t^*_k} q_{\mathrm{r},0,i}^v} = \boldsymbol{B} \boldsymbol{r}_i \, \cancel{\mathrm{e}^{\sigma_i t^*_k} q_{\mathrm{r},0,i}^v} \nonumber \\[0.2em]
	&\Leftrightarrow (\boldsymbol{K} + \sigma_i \boldsymbol{D} + \sigma_i^2 \boldsymbol{M}) \, \boldsymbol{v}_i = \boldsymbol{B} \, \boldsymbol{r}_i.
	\end{align}
	In this special case, the reduction parameters condensate to $(s_{v_i}, \boldsymbol{r}_{_i}, \cancel{q_{\mathrm{r},0,i}^v, t_k^*})$.
\end{itemize} 


\vspace{-0.5em}
\subsubsection*{b) Computational effort}
The presented reduction technique is \emph{simulation-free} or \emph{simulation-lean}, since it ``only" involves the solution of (at most $r \cdot K$) nonlinear systems of equations (NLSE) of full order dimension $n$. These NLSEs can be solved using a Newton-Raphson scheme (cf. line \ref{al:line:Newton}). 


Reduction techniques like POD require a forward numerical simulation (one/multistep, explicit/implicit, adaptive/fixed step-size) of the FOM to gather the snapshots. In case of an implicit scheme, the computational effort of POD compared to NLMM is supposed to be higher, since -- within an implicit simulation -- a NLSE must be solved in \emph{each} time-step with the Newton-Raphson method. 

\vspace{-0.5em}
\subsubsection*{c) Other aspects}
Initial guesses for a NLSE can be taken (depending on the case) from the solution for a zero signal generator $\boldsymbol{v}_{0,i} \!\leftarrow\! \eqref{eq:LP-ZSG-elem}$, from linearized models $\boldsymbol{v}_{0,i} \!=\! \boldsymbol{K}_{\sigma_i}^{-1}\boldsymbol{B} \boldsymbol{r}_i, \boldsymbol{K} \!\!=\! \left. \partial \boldsymbol{f}(\boldsymbol{q})/\partial \boldsymbol{q} \right|_{\boldsymbol{q}_{\textrm{eq}}}$, or from the solutions at neighbouring shifts $\boldsymbol{v}_{0,i+1} \!\leftarrow\! \boldsymbol{v}_{iK}$ or time-snapshots $\boldsymbol{v}_{0,i,k+1} \!\leftarrow\! \boldsymbol{v}_{ik}$. 


A deflation of the basis $\boldsymbol{V}$ (cf. line \ref{al:line:SVD}) is highly recommended, to truncate redundant column vectors and obtain a full rank, orthogonal matrix. Alternatively, an orthogonalization process via Gram-Schmidt or QR-decomposition can optionally be employed.

\vspace{-0.5em}
\subsection{Analysis, Discussion and Limitations}
\vspace{-0.5em}
This section serves as a \emph{complementary} discussion to the one already presented in \cite{cruz2019practicable}.
\begin{itemize}[leftmargin=*]
	\item[$\ast$] As mentioned in Remark~\ref{rem:LPvsNLP}, a linear projection represents a special case of the most general nonlinear projection or the polynomial expansion-based ansatz \eqref{eq:power_series_projection} proposed in \cite[Ch. 4]{huang2004nonlinear}. Applying e.g. a quadratic/cubic projection (see e.g.~\cite{rutzmoser2018model}) or a series expansion ansatz with basis functions customized for the nonlinear system at hand (\cite{scarciotti2017data}) could be superior and even indispensable in certain cases. However, from a numerical perspective, nonlinear projections are much more difficult to handle than linear ones. Moreover, the latter are often used and might be sufficient for many types of nonlinearities.
	
	\item[$\ast$] The choice of the signal generator determines (1) the ansatz for the dynamics $\boldsymbol{q}_{\mathrm{r}}^v(t), \dot{\boldsymbol{q}}_{\mathrm{r}}^v(t), \ddot{\boldsymbol{q}}_{\mathrm{r}}^v(t)$ and (2) the exciting input of the system. Thus, the signal generator should be chosen such that (1) $\boldsymbol{q}_{\mathrm{r}}^v(t)$ constitutes \emph{representative} eigen-/ansatz functions of the underlying nonlinear system and (2) $\boldsymbol{F}(t)$ represents a typical operating input which excites the important dynamics.\\
	Although the validity of a linear signal generator for characterizing nonlinear systems is questionable, this type of signal generator (where complex exponentials serve as ansatz functions) is being implicitly used for the reduction of bilinear and quadratic bilinear systems. Complex exponentials are also being employed in the Fourier approximation ansatz of the Harmonic Balance.\\
	Please note that the linear signal generator \eqref{eq:lin-SG} constitutes a special case of the expansion-based ansatz
	\begin{small}
		\begin{align} \label{eq:power_series_SG}
		\dot{\boldsymbol{q}}_{\mathrm{r}}^v &= \sum_{k=1}^{N} \boldsymbol{S}_v^{(k)} \, {\boldsymbol{q}_{\mathrm{r}}^{v}}^{(k)} = \boldsymbol{S}_{v}^{(1)} \boldsymbol{q}_{\mathrm{r}}^v + \boldsymbol{S}_{v}^{(2)} (\boldsymbol{q}_{\mathrm{r}}^v \otimes \boldsymbol{q}_{\mathrm{r}}^v) + \cdots, \nonumber \\
		\boldsymbol{F} &= \sum_{k=1}^{N} \boldsymbol{R}^{(v)} {\boldsymbol{q}_{\mathrm{r}}^v}^{(k)} = \boldsymbol{R}^{(1)} \boldsymbol{q}_{\mathrm{r}}^v + \boldsymbol{R}^{(2)} (\boldsymbol{q}_{\mathrm{r}}^v \otimes \boldsymbol{q}_{\mathrm{r}}^v) + \cdots,
		\end{align}
	\end{small}
	with $\boldsymbol{S}_v^{(k)} \in \mathbb{C}^{r \times r^k}$, $\boldsymbol{R}^{(k)} \in \mathbb{C}^{m \times r^k}$ and ${\boldsymbol{q}_{\mathrm{r}}^{v}}^{(k)} \in \mathbb{R}^{r^k}$.
	
	

	\item[$\ast$] In the Sylvester-like PDE \eqref{eq:nonlin-time-dep-Sylv-v} the state vector $\boldsymbol{q}_{\mathrm{r}}^v(t)$ cannot be factored out so easily than in \eqref{eq:derivation-Syl-2}. In fact, the key to obtain a \emph{constant}/\emph{state-independent} matrix equation of dimension $n \!\times\! r$ from \eqref{eq:nonlin-time-dep-Sylv-v} lies on both the choice of an adequate projection ansatz (e.g. \eqref{eq:power_series_projection}) and an appropriate signal generator (e.g. \eqref{eq:power_series_SG}), tailored for the nonlinear system at hand. Interestingly, the state-independent matrix Sylvester equations used in bilinear/quadratic bilinear MOR can indeed be derived from \eqref{eq:nonlin-time-dep-Sylv-v} by using the Volterra series representation, with a linear projection and a linear signal generator (\cite{cruz2018nonlinear}). 
	
	
	\item[$\ast$] If a linear projection is applied and the factorization of $\boldsymbol{q}_{\mathrm{r}}^v(t)$ does not succeed, then, lamentably, the \emph{underdetermined} system \eqref{eq:LP-NSG} is obtained. The proposed column-wise consideration has the limitation that \eqref{eq:LP-NSG} is generally not fulfilled, since the couplings in $\boldsymbol{V}\boldsymbol{q}_{\mathrm{r}}^v(t)$, $\boldsymbol{V}\boldsymbol{s}_v(\boldsymbol{q}_{\mathrm{r}}^v(t))$ and $\boldsymbol{r}(\boldsymbol{q}_{\mathrm{r}}^v(t))$ are not being considered. We are currently working on possible ways to numerically solve the underdetermined systems \eqref{eq:LP-NSG}, \eqref{eq:LP-LSG} and \eqref{eq:LP-ZSG}.
	
	 
	\item[$\ast$] In \cite{astolfi2010model}, the signal generator is assumed to be Poisson stable\footnote{Corresponds to $\uplambda(\boldsymbol{S}_v) \in \mathbb{C}^0$ in the linear setting, i.e. exciting the system with a permanent oscillation (\cite[Ch. 8]{isidori1995nonlinear}).}, so that the steady-state of the nonlinear system is well-defined. In linear MOR, however, shifts $\sigma_i \in \mathbb{C} \setminus \uplambda^2(\boldsymbol{M}, \boldsymbol{D}, \boldsymbol{K})$ can be used. Indeed, due to the Meier-Luenberger conditions, the shifts are often chosen on the right half-plane ($\sigma_i \in \mathbb{C}^{>0}$), meaning that the system is being excited by \emph{growing} exponentials. We believe that, within the NLMM algorithm, the nonlinear system can be excited by a stable, Poisson stable or even an unstable generator, as long as the considered time interval for $t_k^*$ is (naturally) well chosen, but \emph{limited}.    
	
	
\end{itemize}

\section{Numerical Example} 
\vspace{-0.5em}
The SO-NLMM algorithm is illustrated by means of a cantilever beam. The 2D model is discretized using 246 triangular elements with quadratic shape functions, yielding (after Dirichlet boundary conditions) $n\!=\!1224$ degrees of freedom in $x$- and $y$-direction. The beam is made of steel, which is modeled as \emph{linear} Kirchhoff \emph{material} with $E\!=\!\unit[210]{GPa}$, $\nu \!=\! \unit{0.3}$ and $\rho \!=\! \unit[1 \cdot 10^4]{\nicefrac{kg}{m^3}}$. However, the model exhibits \emph{geometric nonlinear} behavior due to the used Green-Lagrange strain tensor, which is a \emph{quadratic} function of the nodal displacements. The model equation is given by \eqref{eq:nonlin-FOM}, where zero damping ($\boldsymbol{D}\!=\!\mathbf{0}$) is assumed. The input force $F(t)$ is applied at the tip in negative $y$-direction. The output $y(t)$ is the $y$-displacement of the tip.\\
We use the open-source research code AMfe (\cite{rutzmoser2018model}) for the setup and numerical simulation of the finite element model. Gmsh and ParaView serve hereby as mesh generation and post-processing tools, respectively. The SO-NLMM algorithm \ref{alg:nlmm} has been implemented in Python using a self-programmed Newton-Raphson scheme. 

\begin{figure}[htp!]
	\centering
	\includegraphics[width=0.45\textwidth]{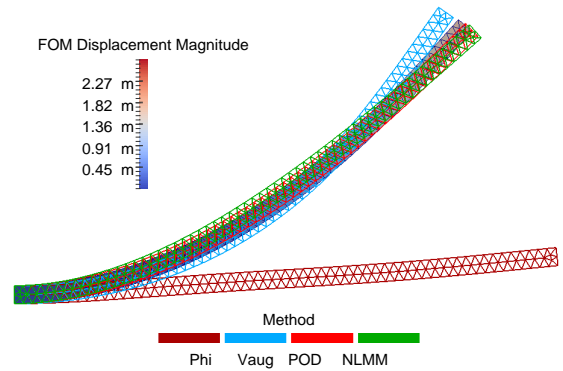}
	\caption{\footnotesize Displacement snapshot of the cantilever beam ($r_{\boldsymbol{\phi}} \!=\! 3,\\
		r_{\text{aug}}\!=\!9, r\!=\!10$).}
	\label{fig:displ-snapshot}
\end{figure}

\begin{figure}[h!]
	\begin{center} 
		\setlength\mywidth{0.39\textwidth}
		\setlength\myheight{0.5\mywidth} 
			\input{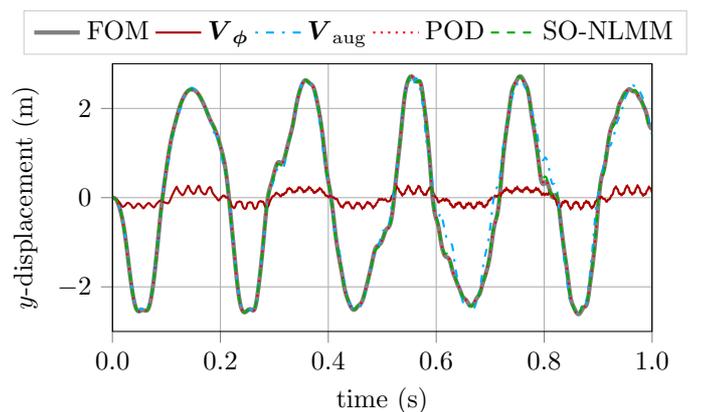}
	\end{center}
	\vspace{-1em}
	\caption{\small $y$-displacement of beam's tip for FOM and different ROMs ($r_{\boldsymbol{\phi}} \!=\! 3,		r_{\text{aug}}\!=\!9, r\!=\!10$).}
	\label{fig:y_displacements}
\end{figure}

\begin{figure}[h!]
	\begin{center} 
		\setlength\mywidth{0.39\textwidth}
		\setlength\myheight{0.5\mywidth} 
			\input{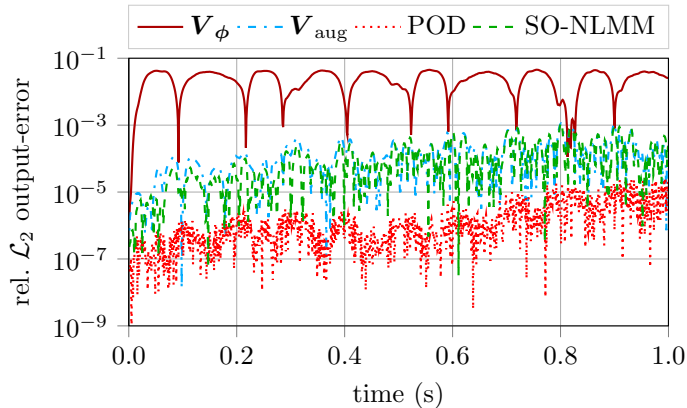}
	\end{center}
	\vspace{-1em}
	\caption{\small $\frac{|y(t) - y_{\mathrm{r},\{*\}}(t)|}{||y(t)||_{\mathcal{L}_2}}$-error for different ROMs $\{*\}$ ($r_{\boldsymbol{\phi}} \!=\! 5$, $r\!=\!20$).}
	\label{fig:error_output_rel}
\end{figure}

We apply Algorithm \ref{alg:nlmm} using a single signal generator with $K \!=\! 10$ or $K \!=\! 20$  equidistant time-snapshots in the interval $\left[0, 1\right]$. For the \emph{training phase} of SO-NLMM and POD, we use the signal generator $q_{\mathrm{r}}^v(t)\!=\! \sin(10 t)$ with corresponding $\dot{q}_{\mathrm{r}}^v(t)$, $\ddot{q}_{\mathrm{r}}^v(t)$ and the training input $F(t) \!=\! r(q_{\mathrm{r}}^v(t)) \!=\! 10^8 \cdot q_{\mathrm{r}}^v(t)$. For the \emph{test phase} of FOM and ROMs, we apply the different input $F(t) \!=\! 10^8 \sin(31 t)$. We compare both approaches with ROMs obtained via $\boldsymbol{V}_{\boldsymbol{\phi}}$ containing only linear vibration modes $\boldsymbol{\phi}_i$, and an augmented basis $\boldsymbol{V}_{\text{aug}}$ containing vibration modes and static modal derivatives to capture the nonlinear behavior. The numerical integration of FOM and ROMs is accomplished by an \emph{implicit} generalized-$\alpha$ scheme with fixed step-size $h\!=\!\unit[1]{ms}$. In terms of approximation quality, some numerical results are given in Fig. \ref{fig:displ-snapshot}, \ref{fig:y_displacements} and \ref{fig:error_output_rel}. The pure linear basis $\boldsymbol{V}_{\boldsymbol{\phi}}$ cannot capture the nonlinear behavior at all. Both SO-NLMM and POD yield satisfactory results, being POD superior. In terms of computational effort, POD required $\approx\!\unit[92]{s}$, whereas SO-NLMM needed only $\approx\!\unit[13]{s} \ (r\!=\!10)$ or $\approx\!\unit[19]{s} \ (r\!=\!20)$ to compute the reduction basis.
\vspace{-0.8em}
\section*{Source Code}
\vspace{-0.8em}
The Python implementation of SO-NLMM and further numerical results including some videos are available at \url{https://zenodo.org/record/2611120}.



\newpage
\bibliographystyle{abbrv}
\bibliography{NLMOR_SONLMM}
\vspace{-0.5em}
\end{document}